\documentclass[aps,prl,superscriptaddress,twocolumn,10pt,a4paper,nofootinbib,showkeys,tightenlines]{revtex4}
\usepackage[left=2cm,right=2cm,top=2cm,bottom=2cm]{geometry}

\usepackage[english]{babel}
\usepackage[latin1]{inputenc}
\usepackage{enumerate}
\usepackage{amsmath}
\usepackage{amsthm}
\usepackage{amssymb}
\usepackage{graphicx}
\usepackage{floatflt}
\usepackage{fancybox}
\usepackage{appendix}
\usepackage{enumitem}
\usepackage{subfigure}
\usepackage{array}
\usepackage{blkarray}

\usepackage[colorlinks,allcolors=blue]{hyperref}
\usepackage[b]{esvect}

\usepackage[normalem]{ulem}
\usepackage{color}

\renewcommand{\Im}{\textrm{Im}}

\newcommand{\om}{\omega}

\newcommand{\lam}{\lambda}
\newcommand{\Gam}{\Gamma}

\newcommand{\varep}{\varepsilon}

\newcommand{\M}{M}

\newcommand{\p}{\partial}
\newcommand{\id}{\mathbf{1}} 
\renewcommand{\d}{\mathrm{d}}

\renewcommand{\leq}{\leqslant}
\renewcommand{\geq}{\geqslant}

\newcommand{\be} {\begin{equation}}
\newcommand{\ee} {\end{equation}}
\newcommand{\bsub}{\begin{subequations}}
\newcommand{\esub}{\end{subequations}}
\newcommand{\bea}{\begin{eqnarray}}
\newcommand{\eea}{\end{eqnarray}}
\newcommand{\bi} {\begin{itemize}}
\newcommand{\ei} {\end{itemize}}
\newcommand{\ben} {\begin{enumerate}}
\newcommand{\een} {\end{enumerate}}
\newcommand{\bmat} {\begin{pmatrix}}
\newcommand{\emat} {\end{pmatrix}}
\newcommand{\bal} {\begin{aligned}}
\newcommand{\eal} {\end{aligned}}
\newcommand{\btab}{\begin{tabular}}
\newcommand{\etab}{\end{tabular}}

\newcommand{\eq}[1]{equation~\eqref{#1}}

\begin{document}
\selectlanguage{english}

\title{Topologically invisible defects in chiral mirror lattices}

\author{Antonin Coutant}
\email{coutant@lma.cnrs-mrs.fr}
\affiliation{Aix Marseille Univ., CNRS, Centrale Marseille, LMA UMR 7031, Marseille, France}
\affiliation{Laboratoire d'Acoustique de l'Université du Mans (LAUM), UMR 6613, Institut d'Acoustique - Graduate School (IA-GS), CNRS, Avenue O. Messiaen, F-72085 Le Mans Cedex 9, France}
\affiliation{Institut de Math\' ematiques de Bourgogne (IMB), UMR 5584, CNRS, Universit\' e de Bourgogne Franche-Comt\' e, F-21000 Dijon, France}

\author{Li-Yang Zheng}
\email{zhengly27@mail.sysu.edu.cn}
\affiliation{School of Science, Shenzhen Campus of Sun Yat-sen University, Shenzhen, China}
\affiliation{Laboratoire d'Acoustique de l'Université du Mans (LAUM), UMR 6613, Institut d'Acoustique - Graduate School (IA-GS), CNRS, Avenue O. Messiaen, F-72085 Le Mans Cedex 9, France}

\author{Vassos Achilleos} 
\email{achilleos.vassos@univ-lemans.fr}
\affiliation{Laboratoire d'Acoustique de l'Université du Mans (LAUM), UMR 6613, Institut d'Acoustique - Graduate School (IA-GS), CNRS, Avenue O. Messiaen, F-72085 Le Mans Cedex 9, France}

\author{Olivier Richoux}
\email{olivier.richoux@univ-lemans.fr}
\affiliation{Laboratoire d'Acoustique de l'Université du Mans (LAUM), UMR 6613, Institut d'Acoustique - Graduate School (IA-GS), CNRS, Avenue O. Messiaen, F-72085 Le Mans Cedex 9, France}

\author{Georgios Theocharis}
\email{georgios.theocharis@univ-lemans.fr}
\affiliation{Laboratoire d'Acoustique de l'Université du Mans (LAUM), UMR 6613, Institut d'Acoustique - Graduate School (IA-GS), CNRS, Avenue O. Messiaen, F-72085 Le Mans Cedex 9, France}

\author{Vincent Pagneux}
\email{vincent.pagneux@univ-lemans.fr}
\affiliation{Laboratoire d'Acoustique de l'Université du Mans (LAUM), UMR 6613, Institut d'Acoustique - Graduate School (IA-GS), CNRS, Avenue O. Messiaen, F-72085 Le Mans Cedex 9, France}

\date{\today}

\begin{abstract}
	One of the hallmark of topological insulators is having conductivity properties that are unaffected by the possible presence of defects. In this work, we go beyond backscattering immunity and obtain topological invisibility across defects or disorder. 	Using a combination of chiral and mirror symmetry, the transmission coefficient is guaranteed to be unity. Importantly, but no phase shift is induced making the defect completely invisible. Many lattices possess the chiral-mirror symmetry, and we choose to demonstrate the principle on an hexagonal lattice model with Kekul\' e distortion displaying topological edge waves, and we show analytically and numerically that the transmission across symmetry preserving defects is unity. We then realize this lattice in an acoustic system, and confirm the invisibility with numerical experiments. We foresee that the versatility of our model will trigger new experiments to observe topological invisibility in various wave systems, such as photonics, cold atoms or elastic waves. 
\end{abstract}


\maketitle


%
%
\section{Introduction}

One of the most appealing property of topological insulators, is that they host edge states that are immune to backscattering. For this reason, topological concepts have attracted a lot of attention in the realm of classical waves~\cite{Ozawa19,Ma19,ZangenehNejad20,Yves22} such as photonics, mechanics or acoustics, as a way to efficiently transport and guide wave energy. In particular, time-reversal topological insulators offers the possibility of immunity to backscattering with purely passive materials. 

To achieve this with classical waves, two main routes have been followed. The first one consists in emulating the quantum spin Hall effect~\cite{Wu15,Suesstrunk15,Mousavi15,Wu16,He16,Yves17,Liu17c,Liu19,Deng20,Sun23}. While in condensed matter the backscattering immunity is guaranteed by Kramers degeneracy of half spin particles, in the classical context this property is not available. In fact, the effective spin (or pseudo-spin) usually relies on crystal symmetries, which are broken as soon as defects are introduced. The second route is to start with a material with a pair of inequivalent Dirac points, where the valley polarization plays a role similar to the electronic spin (valley Hall effect)~\cite{Lu17,Pal17,Chen18,Lu18,Laforge19,Tian20,Zheng20}. However, valley conservation is only approximate when periodicity is broken. Due to the above limitations, recent works have shown that backscattering is unavoidable in these systems~\cite{Orazbayev19,Arregui21,Rosiek23,Rechtsman23}.

In this work, we show that it is possible to obtain not only immunity to backscattering but also perfect invisibility in two-dimensional (2D) passive lattices using both chiral and mirror symmetries. This combination of symmetries allows for topological phases characterized by nontrivial mirror winding numbers. Here we focus on the case of hexagonal lattices with Kekulé distortion~\cite{Wu16,Liu17b,Kariyado17,Xie19,Liu19,Liu19b,Freeney20,Zhang20,Wakao20}. We show that if the system has commuting chiral and mirror symmetries, and there is a single pair of propagating waves near zero energy, then the transmission coefficient is either zero or unity across symmetry preserving defects or disorder. This surprising and important property has never been reported in topological materials. Additionally, if appropriately defined topological indices of the defect are trivial, then transmission is guaranteed to be one. This is illustrated in Fig.~\ref{Principle_Fig}. 

Remarkably, the transmission coefficient across topologically trivial defects not only has a unit modulus but also a vanishing phase, which means that the defect is completely invisible. This result can be applied to a plethora of classical waves systems in various ways such as a tight binding approximation of a set of coupled resonators~\cite{Freeney20,Zhang20} or mass and springs systems~\cite{Wakao20}. We show that this model can be realized with a network of acoustic waveguides, and the relevant symmetries obtained to a very high degree of accuracy. Topological invisibility of symmetry preserving defects for acoustic edge waves is confirmed by numerical simulations of the Helmholtz equation.

%
%
\section{Chiral mirror models}

\begin{figure*}[htp]
	\centering
	\includegraphics[width=\textwidth]{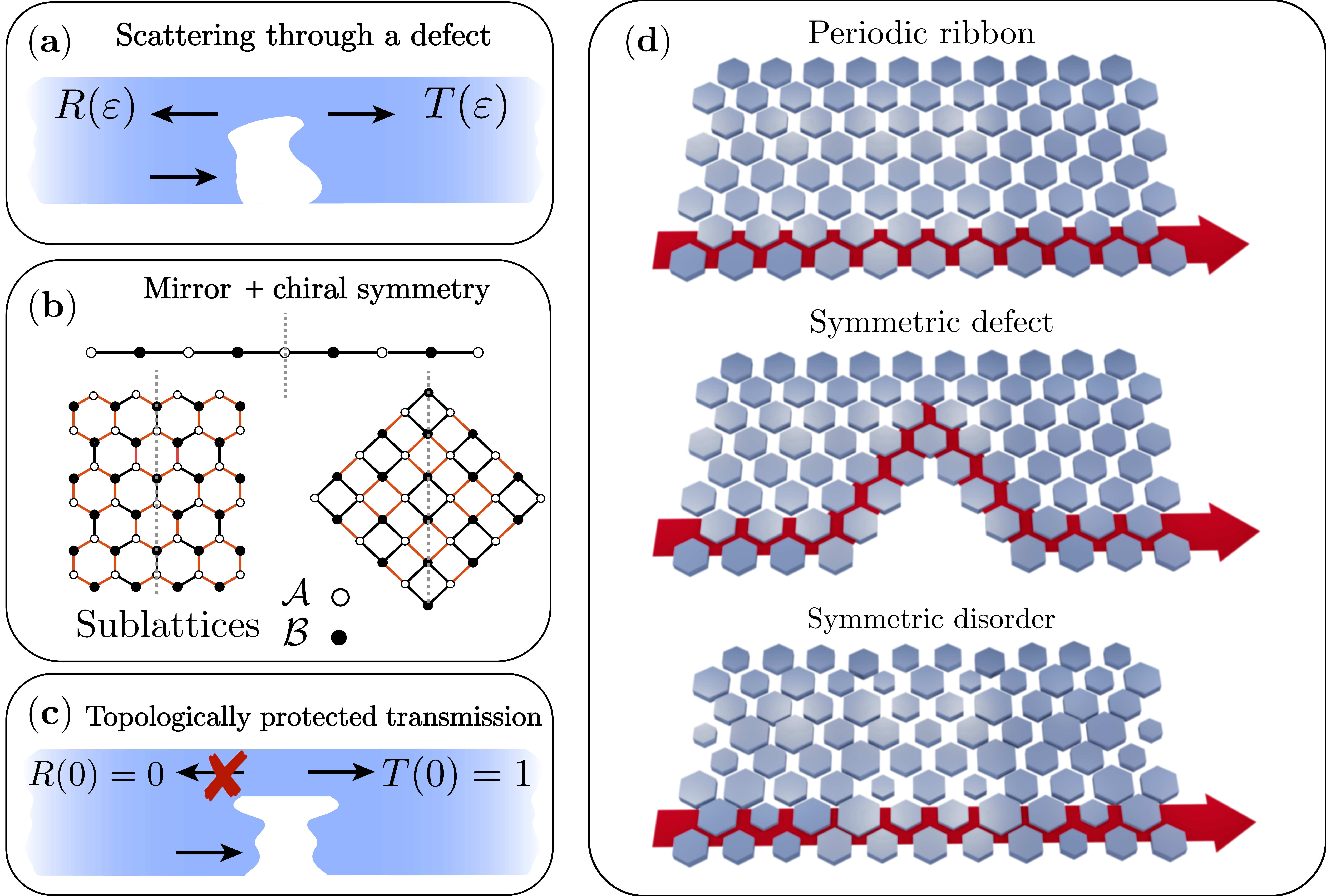}
	\caption{Principle of topological invisibility with chiral and mirror symmetries. (a) Illustration of the scattering problem in a ribbon with a defect. (b) Examples of unit cells of models displaying commuting chiral and mirror symmetries. Sites of sublattice $\mathcal A$ (resp. $\mathcal B$) are hollow circles (resp. plain circles), and the axis of mirror symmetry is shown as a grey dashed line. (c) Illustration of perfect transmission in a chiral and mirror symmetric ribbon. (d) Illustrations of ribbons with unit transmission. 
	}
	\label{Principle_Fig} 
\end{figure*}

We consider a lattice model described by a hermitian Hamiltonian of the form $H = \sum_{i, j} t_{i,j} \hat a_i^\dagger \hat a_j$, where $\hat a_j$ is the annihilation operator on site $j$ and $t_{i,j} = t_{j, i}$ are real hopping coefficients, and we look for wave functions $\Phi$ solution of $H \Phi = \varep \Phi$ where $\varep$ is the energy. The key assumption in this work is that the model has two commuting symmetries: \emph{chiral} and \emph{mirror symmetry}. In Fig.~\ref{Principle_Fig}-(b), we show several examples of such models.

Chiral symmetry, or sublattice symmetry means that the lattice can be decomposed into two sublattices $\mathcal A$ and $\mathcal B$ such that hoppings only connect sites of different sublattices. Algebraically, this can be translated by introducing the chiral operator 
\be \label{Chiral_Op}
\Gam = \mathrm{diag}(- \mathbf{1}_{\mathcal A}, \mathbf{1}_{\mathcal B}) , 
\ee 
i.e. it flips the sign of the amplitudes on sublattice $\mathcal A$ while leaving the amplitudes of sublattice $\mathcal B$ unchanged. Chiral symmetry is equivalent to the anti-commutation relation 
\be \label{Chiral_Sym}
\Gam H + H \Gam = 0. 
\ee 
The model is also considered mirror symmetric, which means there is a mirror operator $\M$ that commutes with the Hamiltonian: 
\be \label{Mirror_Sym}
\M H - H \M = 0 . 
\ee 
Moreover, we assume that both operators $\Gam$ and $\M$ commute: 
\be \label{Chirror_Sym}
\M \Gam - \Gam \M = 0 . 
\ee 
In other words, mirror symmetry respect the sublattice structure: the mirror of a site in $\mathcal A$ (resp. $\mathcal B$) is also in $\mathcal A$ (resp. $\mathcal B$), as shown in Fig.~\ref{Principle_Fig}-(b).

\section{Edge waves in the Kekule model}

\begin{figure*}[htp]
	\centering
	\includegraphics[width=\textwidth]{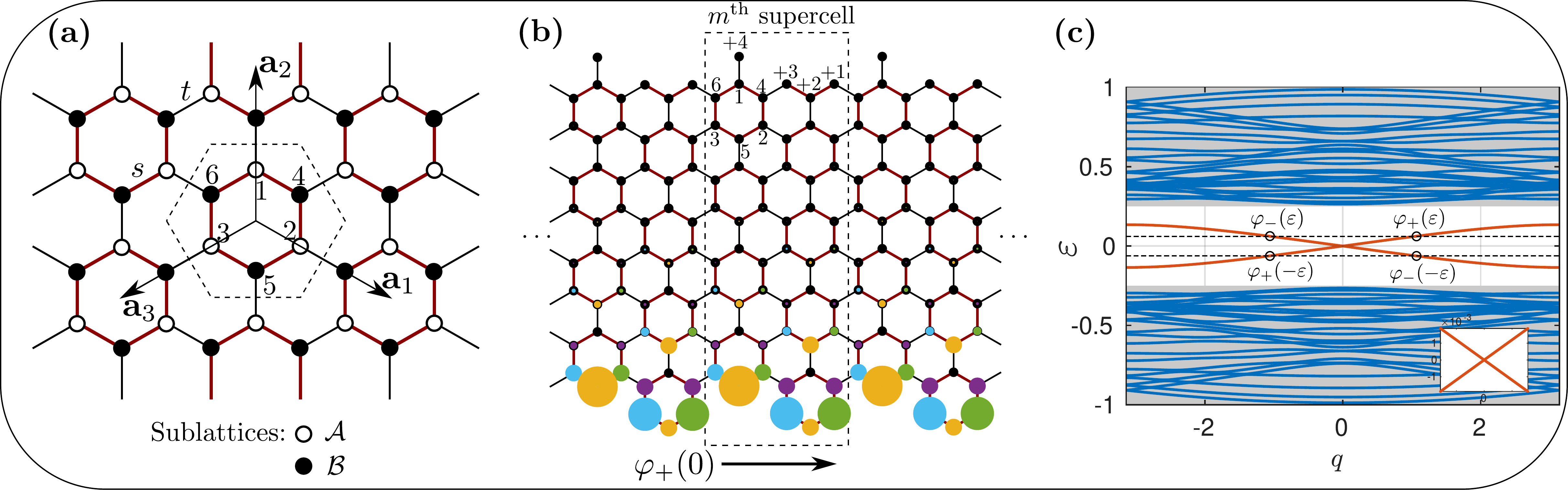}
	\caption{Main properties of Kekulé ribbons with mixed edges. (a) Bulk structure of the Kekule model. Sites of sublattice $\mathcal A$ (resp. $\mathcal B$) are hollow circles (resp. plain circles), and intracell indices $1..6$ are indicated. (b) Ribbon with mixed edges: molecular zigzag down and partially bearded up. Numbers with the $+$ sign label the extra sites compared to a canonical molecular zigzag ribbon. Colored circles show the zero energy edge mode profile inside a ribbon supercell ($q \to 0^+$ and $\varep \to 0^+$). Circle radii give the field absolute value, and colors show the phase: due to the symmetries of the problem, the phase is either $0$ (yellow), $\pi/2$ (green), $\pi$ (purple) or $-\pi/2$ (cyan). We took $N_y=4$, $s=0.25$, $t=0.5$.  (c) Dispersion relation of modes in a mixed edge ribbon for $N_y=4$, $s=0.25$, $t=0.5$. Edge modes are emphasized in red and bulk bands are shaded. 
	}
	\label{Kekule_Fig} 
\end{figure*}

Among the many models possessing properties \eqref{Chiral_Sym}, \eqref{Mirror_Sym}, \eqref{Chirror_Sym}, we also need a single pair of propagating waves near zero energy. Hence, we now focus on the Kekulé model, which has been shown to possess a pair of topologically protected edge modes~\cite{Wu16,Kariyado17,Liu17b,Liu19,Xie19}. The model is made of a honeycomb lattice with nearest neighbour hoppings. A Kekulé distortion is added by defining hexagonal molecules of 6 sites with different intracell hoppings $t_{i,j} = s$ and extracell ones $t_{i,j} = t$. This is illustrated in Fig.~\ref{Kekule_Fig}-(a). The distortion opens a gap around zero for energies $|\varep | < |t-s|$. The model is chiral and possesses mirror symmetries whose reflection planes passe through sites of a molecule (i.e. along $\mathbf{a}_{j=1..3}$ in Fig.~\ref{Kekule_Fig}-(a)) commute with $\Gam$. 

The combination of chiral and mirror symmetry allows for the construction of topological invariants: the mirror winding numbers. These topological invariants for the Kekulé model with different boundary types have been studied in details in~\cite{Kariyado17}. In particular, molecular zigzag and partially bearded edges are relevant for us because they preserve an appropriate mirror symmetry and hence, can host topological edge modes. The authors of~\cite{Kariyado17} showed that the former is topological when $s<t$ and trivial when $s>t$, while the latter is trivial for $s<t$ and topological for $s>t$. For half-space configurations (only one edge) in a topological phase, these edge modes are gapless: there is a Dirac point at $\varep = 0$. However, in ribbon configurations, the finite width may open a minigap around $\varep = 0$ due to the interaction between the lower and upper edge~\cite{Wu16,Liu17}. 

Since we need a single pair of propagating waves at $\varep = 0$, we now show how to construct ribbons displaying edge waves with a vanishing minigap. The width $N_y$ is defined as the number of molecules vertically aligned, for example $N_y=4$ in Fig.~\ref{Kekule_Fig}(b). Modes of the ribbon are obtained by solving the eigenvalue problem 
\be \label{WG_Ham}
\varep \phi = H_{\rm rib}(q) \phi, 
\ee 
with $H_{\rm rib}$ the Bloch Hamiltonian of a ribbon supercell and $q$ the dimensionless Bloch momentum in the longitudinal direction ($-\pi<q<\pi$). To avoid the opening of a minigap, we choose different types of edges on the lower and upper side: a molecular zigzag edge and a partially bearded edge, as in Fig.~\ref{Kekule_Fig}-(b). A ribbon constructed this way has always a single pair of edge waves: if $s<t$ it is localized along the molecular zigzag edge, and if $s>t$ it is localized along the partially bearded one. Without loss of generality, we focus on the former case ($s<t$). The dispersion relation of the modes is shown in Fig.~\ref{Kekule_Fig}-(c), and the profile of the edge mode at zero energy in Fig.~\ref{Kekule_Fig}-(b). 

As we now show, in this type of ribbons the existence of an exact Dirac point at zero energy is guaranteed by symmetry. To demonstrate this, we apply a method developed in~\cite{Koshino14} to the ribbon of Fig.~\ref{Kekule_Fig}-(b), with which it is possible to guarantee the existence of zero energy modes using a combination of spatial symmetry and chiral symmetry. The main result needed, sometimes referred to as Lieb theorem~\cite{Sutherland86,Lieb89,Inui94}, is that if a finite chiral lattice has a number $N_1$ of sites on one sublattice larger than the number of sites $N_2$ on the other sublattice, then, there is at least $N_1 - N_2$ zero energy solutions with support on the first sublattice. The idea is to apply this to a ribbon supercell, or more precisely, to $H_{\rm rib}(q)$. Such a supercell contains an equal number of sites on $\mathcal A$ and $\mathcal B$, so zero modes are not guaranteed by chiral symmetry alone. However, at $q=0$, $H_{\rm rib}$ is also mirror symmetric (both equations \eqref{Chiral_Sym} and \eqref{Mirror_Sym} hold). Hence, we first block diagonalize $H_{\rm rib}$ into mirror symmetric and mirror anti-symmetric subspaces. Each subspace has now an uneven number of basis vectors on the two sublattices, which allows us to use the previous result. Explicitly, the symmetric ribbon Hamiltonian has one extra basis vector on $\mathcal B$, while the anti-symmetric ribbon Hamiltonian has one extra basis vector on $\mathcal A$ (a detailed counting is provided in Supplemental material). This guarantees the existence of two solutions at $q=0$ and $\varep=0$ with support on each sublattice, and corresponding to the Dirac point of edge waves. This is emphasized by the inset of Fig.~\ref{Kekule_Fig}-(c), which is a zoom near the Dirac point. Moreover, the above counting also gives us on which sublattice each mode lives: the symmetric mode $\varphi_S$ has support on $\mathcal B$, while the anti-symmetric one $\varphi_A$ has support on $\mathcal A$. This can be written 
\be \label{ASmodes_support}
\Gam \varphi_S = \varphi_S, \quad \text{and} \quad \Gam \varphi_A = -\varphi_A .
\ee

%
%
\section{Invisibility of symmetry preserving defects}

We now analyze the propagation properties of these edge waves across defects. Let us directly state our main result: 
\bigskip

\noindent {\bf Theorem:} \emph{If a ribbon (with edges as in Fig.~\ref{Kekule_Fig}-(b)) has a chiral mirror symmetric defect made of missing or added molecules and/or modified hopping values, then that defect is invisible for edge waves at zero energy: } 
\be \label{Theorem_eq}
T(0) = 1. 
\ee 

\begin{figure*}[htp]
	\centering
	\includegraphics[width=\textwidth]{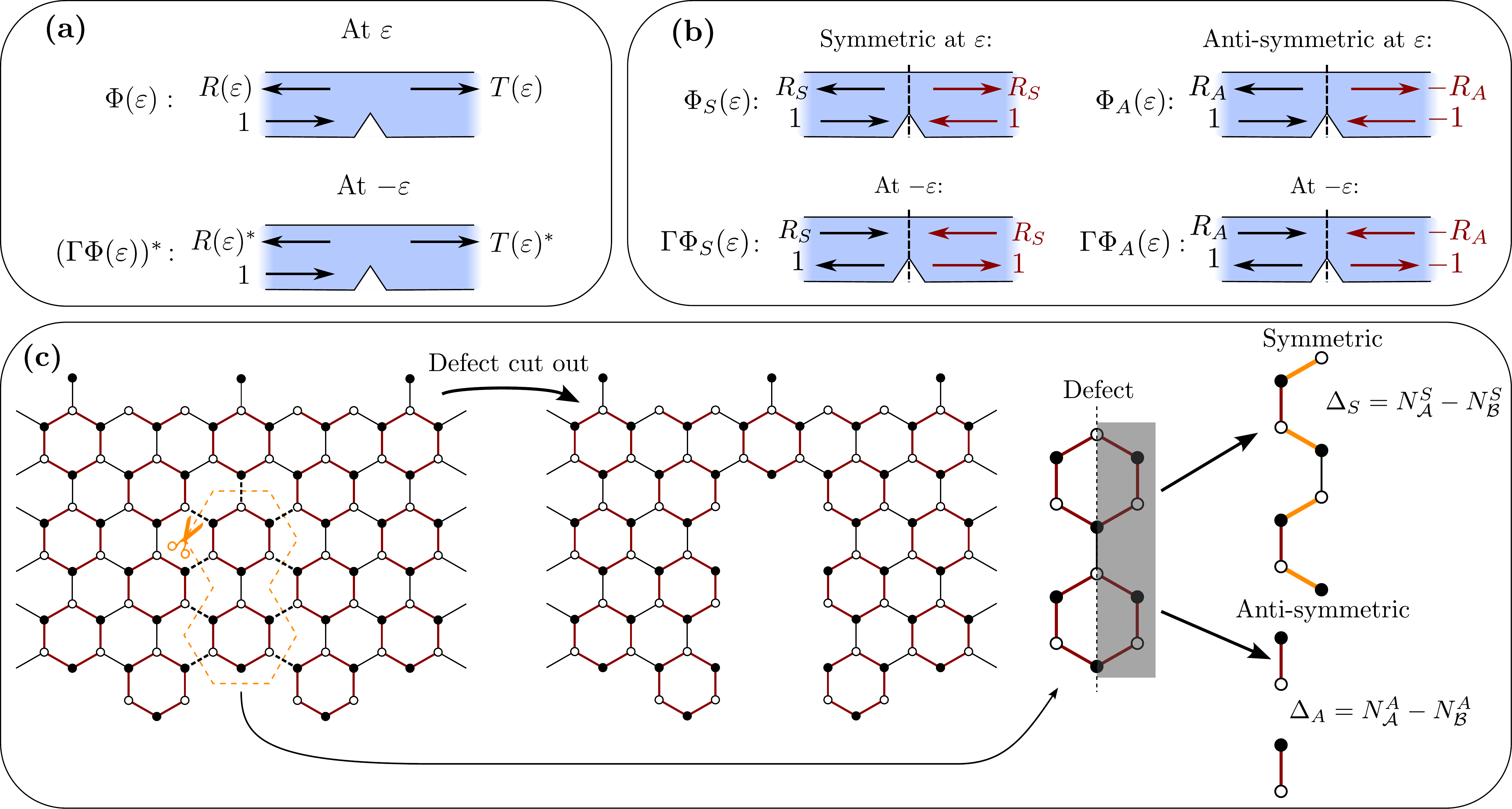}
	\caption{Illustration of symmetry properties of scattering solutions and topological indices of defects. (a) Illustration of chiral symmetry (equations~\eqref{Chiral_Modes} and \eqref{TR_Modes} implying $(\Gam \varphi_\pm(\varep))^* = \varphi_\pm(-\varep)$) applied on a scattering solution. (b) Illustration of mirror symmetry (equation~\eqref{Mirror_Modes}) applied on a scattering solution. (c) Illustration of how to obtain a defect's topological indices, as defined in equation~\eqref{TopoIndices}. In the symmetric part of the defect, some hopping take the value $\sqrt{2} s$ (orange) due to the normalization of the new basis vectors (see Supplemental material). This does not affect the topological indices. 
	}
	\label{Symmetry_Fig} 
\end{figure*}

To show this, we first assume that $\varep$ is around $0$, such that there are only two propagating edge waves (see Fig.~\ref{Kekule_Fig}-(c)), which we call $\varphi_+(\varep)$ and $\varphi_-(\varep)$. The scattering problem is to find a solution of $\varep \Phi = H \Phi$ with $H$ the infinite ribbon Hamiltonian containing the defect, such that asymptotically: 
\bsub \label{Scatt_as} \bea  
\Phi(\varep) &\underset{-\infty}{\sim} & \varphi_+(\varep) + R(\varep) \varphi_-(\varep) , \\
 &\underset{+\infty}{\sim} & T(\varep) \varphi_+(\varep) . 
\eea \esub 
Notice that this definition of the scattering coefficients ensures that $T(\varep)=1$ in the absence of defect. We now analyze the consequence of the symmetries of the problem on the scattering coefficients. First, from the relation \eqref{Chiral_Sym}, the chiral operator $\Gam$ applied on an eigenmode at energy $\varep$ gives an eigenmode at energy $-\varep$ and the same momentum $q$. Moreover, it changes its propagation direction, since the group velocity of a mode is given by $v_g = \frac{\d \varep}{\d q}$ (see Fig.~\ref{Kekule_Fig}-(c)). This implies 
\be \label{Chiral_Modes}
\Gam \varphi_+(\varep) = \varphi_-(-\varep) . 
\ee 
In addition, our system is time reversal invariant (Hamiltonian is real), which means that $H_{\rm rib}(q)^* = H_{\rm rib}(-q)$. This gives the identity 
\be \label{TR_Modes}
\varphi_+(\varep)^* = \varphi_-(\varep). 
\ee 
For equations~\eqref{Chiral_Modes} and \eqref{TR_Modes} to hold, we of course need to fix the phase of each modes. Using \eq{Chiral_Op}, this is done by choosing a phase 0 on a site of the sublattice $\mathcal B$. We can now use \eqref{Chiral_Modes} and \eqref{TR_Modes} on \eqref{Scatt_as} to obtain the scattering solution at $-\varep$. Its asymptotic behavior is: 
\bsub \bea  
(\Gam \Phi(\varep))^* &\underset{-\infty}{\sim} & \varphi_+(-\varep) + R(\varep)^* \varphi_-(-\varep) , \qquad \\
&\underset{+\infty}{\sim} & T(\varep)^* \varphi_+(-\varep) . 
\eea \esub 
This is illustrated in Fig.~\ref{Symmetry_Fig}-(a), and we obtain 
\bsub \bea  
T(\varep)^* &=& T(-\varep) , \\
R(\varep)^* &=& R(-\varep) . \label{Chiral_Coef_R}
\eea \esub 
Let us next use the mirror symmetry of the scattering system. Since mirror symmetry applied on the Bloch Hamiltonian flips the sign of $q$, i.e. $\M H_{\rm rib}(q) \M = H_{\rm rib}(-q)$, we obtain  
\be \label{Mirror_Modes}
\M \varphi_+(\varep) = \varphi_-(\varep). 
\ee 
For this identity to hold we need to fix the phase of the modes to be 0 on a site of the symmetry axis, in addition to the previous constraint that this site lies in $\mathcal B$. More details about phase fixing is provided in Supplemental material. 

Now, because we consider a mirror symmetric defect, the scattering problem also has that symmetry. Hence, as is customary we decompose it into a symmetric subproblem and an anti-symmetric one~\cite{Merkel15}, with each subproblem equivalent to a simple one-port reflection problem. All this defines a symmetric reflection coefficient $R_S$ and an anti-symmetric one $R_A$ with the corresponding scattering solutions $\Phi_S$ and $\Phi_A$: 
\be  \label{SA_Scatt}  
\Phi_{S/A}(\varep) \underset{-\infty}{\sim} \varphi_+(\varep) + R_{S/A}(\varep) \varphi_-(\varep) , 
\ee  
as illustrated in Fig.~\ref{Symmetry_Fig}-(b). The scattering coefficients of the full problem are given by~\cite{Merkel15}: 
\bsub \label{ASrefl_Coef} \bea  
T(\varep) &=& \frac{R_S(\varep) - R_A(\varep)}{2} , \\
R(\varep) &=& \frac{R_S(\varep) + R_A(\varep)}{2} . 
\eea \esub 
Since chiral symmetry commutes with mirror symmetry (equation~\eqref{Chirror_Sym} with the mirror symmetry of the ribbon), each subproblem has the chiral symmetry. This means that $R_S$ and $R_A$ satisfy equation~\eqref{Chiral_Coef_R}, and thus are real-valued at $\varep = 0$. Furthermore, by energy conservation, $R_S$ and $R_A$ have unit modulus, implying that they can only be $\pm 1$ at zero energy. Consequently, from equation~\eqref{ASrefl_Coef} the transmission coefficient $T(0)$ can only be $0$ or $\pm 1$. 

Now, to identify which defects have $T(0)=1$, we need to study in details the sublattice structure of the scattering solutions. We first directly apply the chiral operator to the scattering solutions of \eq{SA_Scatt}, as illustrated in Fig.~\ref{Symmetry_Fig}-(b). Then, by linearity, we deduce $R_S(-\varep) = 1/R_S(\varep)$ and, more importantly:  
\bsub \label{AS_SubSupport} \bea  
\Gam \Phi_S(\varep) &=& R_S(\varep) \Phi_S(-\varep) , \\
\Gam \Phi_A(\varep) &=& R_A(\varep) \Phi_A(-\varep) . 
\eea \esub 
Crucially, at $\varep=0$, $\Phi_S$ and $\Phi_A$ are eigenvectors of the chiral operator $\Gam$, with eigenvalues $R_S$ and $R_A$. Thus, $R_S(0)=1$ (resp. $R_A(0)=1$) is equivalent to $\Phi_S$ (resp. $\Phi_A$) has support on sublattice $\mathcal B$, while if $R_S(0)=-1$ (resp. $R_A(0)=-1$) it has support on $\mathcal A$. 

To complete the proof, it is convenient to introduce the topological indices of a chiral mirror defect~\cite{Teo10,Koshino14} in the following manner. The set of removed (or added) sites makes a finite (chiral mirror) lattice. We split this lattice in a symmetric and an anti-symmetric one. We then count the number of sites on each sublattice, $N_{\mathcal A}$ on $\mathcal A$ and $N_{\mathcal B}$ on $\mathcal B$, and define the pair of topological indices as~\footnote{Sites added rather than removed are counted negatively.}: 
\be \label{TopoIndices}
(\Delta_S, \Delta_A) = (N_{\mathcal A}^S - N_{\mathcal B}^S, N_{\mathcal A}^A - N_{\mathcal B}^A) . 
\ee 
This definition is illustrated in Fig.~\ref{Symmetry_Fig}-(c). We now make the conjecture that for a topologically trivial defect, i.e. $(\Delta_S, \Delta_A) = (0,0)$, the scattering solutions $\Phi_S(0)$ and $\Phi_A(0)$ have support on the same sublattice as without defect (this conjecture is further supported in Supplemental material). As we saw in equation~\eqref{ASmodes_support}, this means that $\Phi_S(0)$ has support on $\mathcal B$ and $\Phi_A(0)$ on $\mathcal A$. Using equation~\eqref{AS_SubSupport} this implies that $R_S(0) = 1$ and $R_A(0)=-1$. Then, we conclude by noticing that a defect made of added or removed molecules is always topologically trivial. Hence, using equations \eqref{AS_SubSupport} and \eqref{ASrefl_Coef}, we conclude that on such defects $T(0)=1$. 

Importantly, according to the above analysis, any disordered slab made of random hoppings preserving chiral and mirror symmetries, will be a topologically trivial defect. As a consequence it  will be invisible to zero energy edge waves. Notice that this is reminiscent to zero index materials~\cite{Liberal17}, with the difference that it does not rely on effective parameters of the materials but rather on topological protection.

%
%
\section{Scattering of edge waves on defects}

\begin{figure*}[htp]
	\centering
	\includegraphics[width=\textwidth]{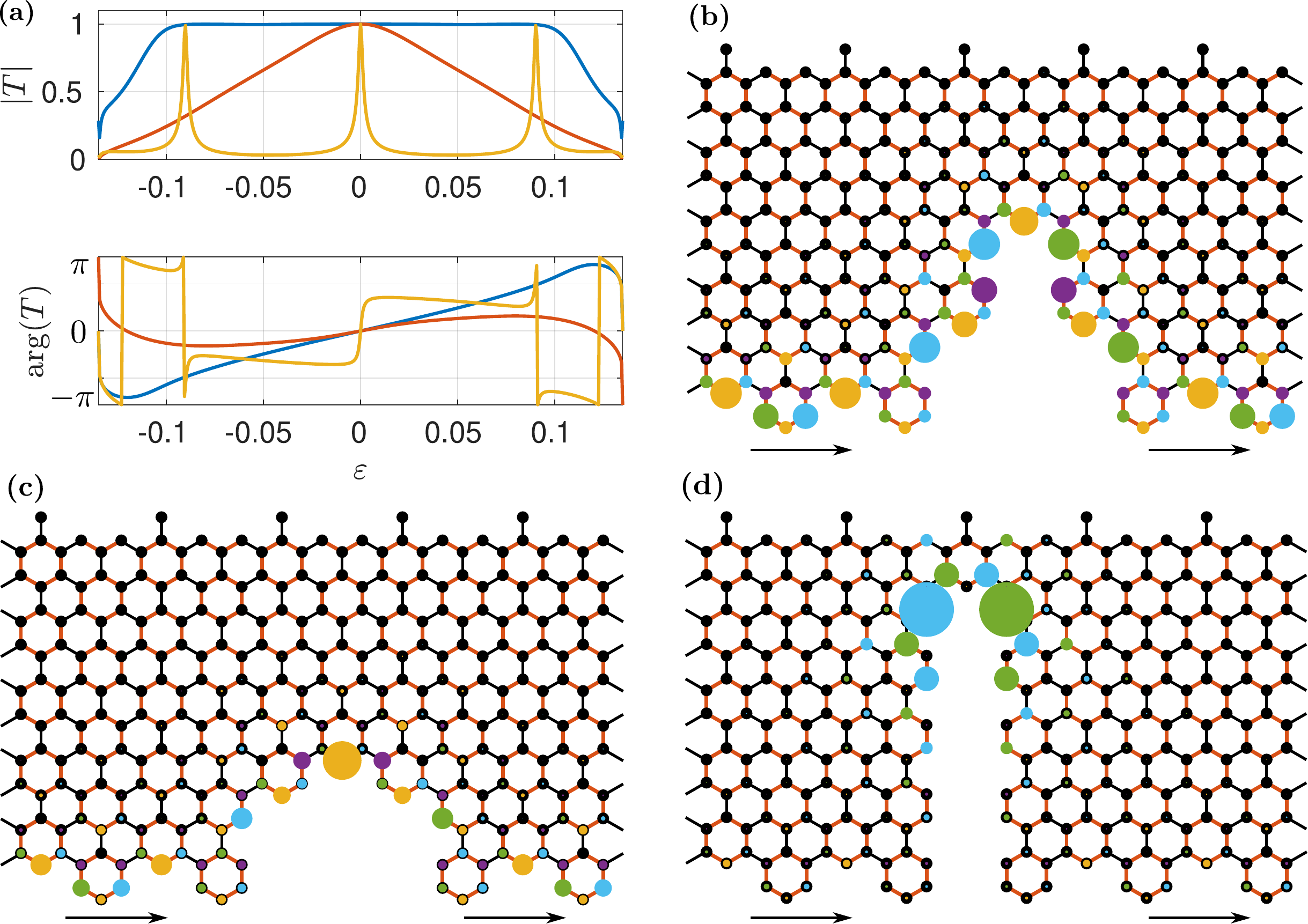}
	\caption{Scattering of edge waves on chiral mirror symmetric defects. (a) Transmission coefficients as a function of the energy of edge waves on the three defects: blue (b), red (c), yellow (d). We took $N_y=5$, $s=0.25$, $t=0.5$ and work with the complex energy $\varep + i \nu$ with $\nu = 10^{-6}$ (see Supplemental material). (b-d) Scattering solutions at $\varep=0$ on the different defects. Circle radii give the field absolute value, and colors show the phase: due to the symmetries of the problem, the phase is either $0$ (yellow), $\pi/2$ (green), $\pi$ (purple) or $-\pi/2$ (cyan). 
	}
	\label{Ribbon_Scatt_Fig} 
\end{figure*}

We now compute the transmission and reflection coefficients over various defects using a transfer matrix formalism adapted from~\cite{Dwivedi16} and presented in Supplemental material. The scattering solution of the ribbon including the defect is obtained as a function of the incident energy $\varep$. In Fig.~\ref{Ribbon_Scatt_Fig}-(a), we show the amplitude and phase of the transmission coefficient for three different chiral mirror symmetric defects. The corresponding defects are illustrated in panels (b)-(d), where the zero energy scattering solutions over these defects are also depicted. As predicted by our theorem, the transmission coefficient is exactly unity (with $|T|=1$ and $\arg(T)=0$ mod $2\pi$) at zero energy for all defects.

At energies around 0, the transmission coefficient behavior varies significantly depending on the shape of the defect, which can be interpreted in the following way. First, for the defect of Fig.~\ref{Ribbon_Scatt_Fig}-(b) the edge waves always follow molecular zigzag boundaries, along which edge waves are propagating near $\varep=0$. Combined with the perfect transmission $T=1$ at $\varep = 0$, one expects a good transmission for a broad range of  energies around 0, which is confirmed by our calculation. On the contrary, if the defect has pieces with edges where edge waves become evanescent, we expect a lower transmission for $\varep$ away from $0$. For instance, the defect of Fig.~\ref{Ribbon_Scatt_Fig}-(c) looks similar to that of (b) but the tip molecule has been put back. As a result, transmission decreases when the energy departs from 0. The defect of Fig.~\ref{Ribbon_Scatt_Fig}-(d) corresponds to a rather extreme case where the edge wave has to propagate along a long armchair boundary. Since edge waves are gapped along armchair boundaries~\cite{Liu17b} we expect an exponentially small transmission for $\varep$ inside that gap. In that respect, it makes the perfect transmission even more surprising, since the preceding line of reasoning would suggest an exponentially small transmission also at $\varep = 0$. In fact, the scattering can be seen as the result of exciting a quasi bound state in the continuum through a tunnel effect, which leads to a unit transmission. We clearly see the resonance with that quasi bound state in Fig.~\ref{Ribbon_Scatt_Fig}-(d). We point out that the broadband phenomenon near zero energy in Fig.~\ref{Ribbon_Scatt_Fig}-(b) is rather close to the valley Hall effect, where edge waves are highly transmitted from one edge to another along the same valley direction. However, in our setup, the additional perfect transmission at $\varep=0$ protected by chiral and mirror symmetry guarantees a higher transmission for similar turns, but also allows edge waves to be transmitted across edges not allowed by valley conservation.

\begin{figure*}[htp]
	\centering
	\includegraphics[width=\textwidth]{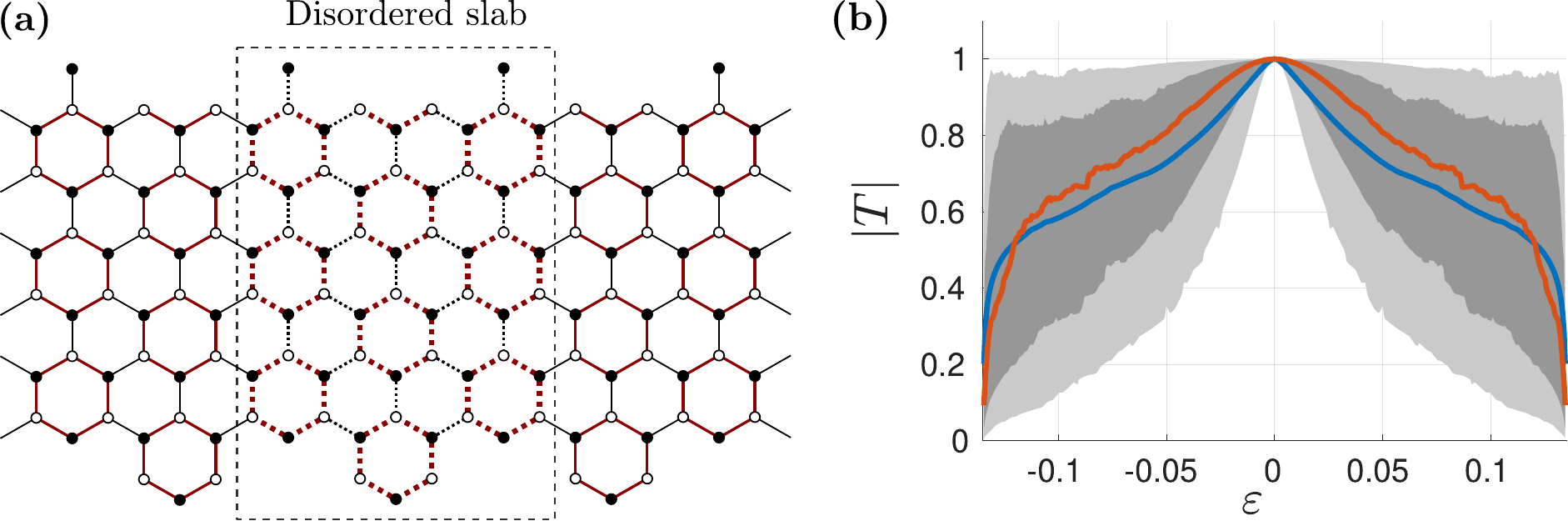}
	\caption{Scattering of edge waves on a disordered slab. (a) Illustration of a symmetric disordered slab of three columns. Hoppings with added random values are marked by dashed lines. (b) Transmission coefficient as a function of edge wave energy on disordered slabs of three columns. We show the mean value $\langle |T|\rangle$ (blue line) and median (red line). The shaded regions mark the first and last quartiles (dark grey) and first and last deciles (light grey). Statistics is taken over 200 realizations. We took $N_y=3$, $s \in [0.05,0.45]$, $t\in [0.3,0.7]$ and work with the complex energy $\varep + i \nu$ with $\nu = 10^{-6}$ (see Supplemental material). 
	}
	\label{Ribbon_ScattDisorder_Fig} 
\end{figure*}

We complete this section by investigating the transmission coefficient trough slabs with mirror and chiral symmetric disorder. The results are summarized in Fig.~\ref{Ribbon_ScattDisorder_Fig}. Inside the slab, we randomly take the hopping coefficients following a uniform probability density around their mean values corresponding to the rest of the ribbon and symmetrize about the central axis (see Fig.~\ref{Ribbon_ScattDisorder_Fig}-(a)). The obtained results confirm our theorem that for all realizations the slab is invisible to the edge waves at $\varep=0$. 

We would like to emphasize that the topological invisibility we are demonstrating, i.e. equation~\eqref{Theorem_eq}, directly comes from mirror and chiral symmetry (more precisely, equations~\eqref{Chiral_Modes} and \eqref{Mirror_Modes}), and only from that. For instance, the same invisibility result is true in 1D simple chains (see Fig.~\ref{Principle_Fig}-(b)) on trivial defects (where topological indices are defined similarly to~\eqref{TopoIndices}). In particular, the results of Fig.~\ref{Ribbon_Scatt_Fig} involve the Kekulé model, in which a notion of pseudo-spin can be defined~\cite{Liu19}, but the pseudo-spin is not at the origin of the topological invisibility. 

\section{Acoustic networks}

We now show that the theoretically predicted invisible defects can be implemented in classical wave systems. To do so we choose a recently proposed acoustic continuous system governed by the Helmholtz equation $\Delta p + k^2 p = 0$ with acoustic frequency $\om = k c_0$ ($c_0$ is the speed of sound) which exactly maps to various discrete lattices and is used to realize various topological phases~\cite{Coutant20,Coutant20b,Coutant21}. The system consists of a network of hollow tubes connected on a graph, here, a honeycomb one. The Kekulé distortion is reproduced by varying the tube cross-sections in the same way as in the lattice model: cross-sections of intracell tubes $\sigma_s$ differ from that of extracell ones $\sigma_t$. We can then show that if all tubes have the same length $L$, and the transverse dimensions are much smaller than $L$, then fixed frequency solutions are obtained as eigenvectors of an effective Hamiltonian on the same graph: $\varep \phi = H \phi$, with $\phi$ a vector containing the pressure values at every node. $\varepsilon$ is the effective energy related to the acoustic frequency by $\varep = \cos(kL)$. Notice that when $0<kL<\pi$, $\varep$ is a decreasing function of $k$ and as a consequence, the acoustic group velocity $v_g^{\rm ac}$ has the sign of $\frac{\d k}{\d q}$, which is opposite to its lattice counterpart $v_g = \frac{\d \varep}{\d q}$. The hopping coefficients between a node $a$ and $b$ are given by the ratio of the tubes cross sections: 
\be 
H_{ab} = \frac{\sigma_{ab}}{\sum_{b'} \sigma_{ab'}} , 
\ee 
where $b'$ are all nodes connected to $a$. To obtain open boundary conditions of the lattice model, we add extra tubes of length $L$ on every nodes on the edge, with an open end enforcing a vanishing pressure condition (see~\cite{Coutant21} for more details). 

\begin{figure*}[htp]
	\centering
	\includegraphics[width=\textwidth]{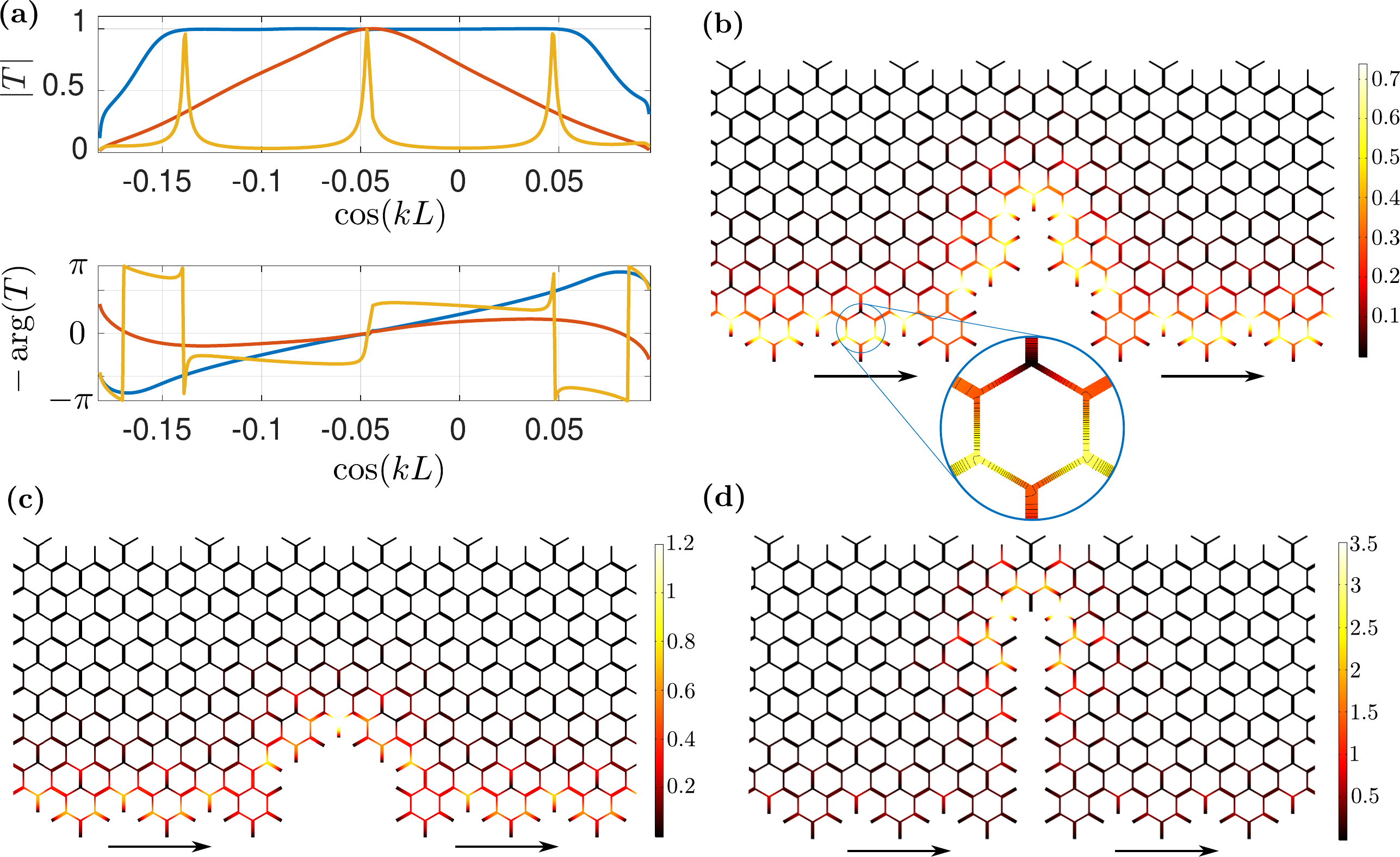}
	\caption{Finite element simulations of the 2D Helmholtz equation in a Kekulé network. We took tubes with $L=40$, $\sigma_s=4$ and $\sigma_t=8$. (a) Transmission coefficients obtained from numerical simulations as a function of $\varep = \cos(kL)$ on the three defects: blue (b), red (c), yellow (d). We took $N_y=5$, $s=0.25$, $t=0.5$. Since the acoustic group velocity sign is flipped with respect to the lattice model, we compare $T^*(\varep)$ with $T(\varep)$ of Fig.~\ref{Ribbon_Scatt_Fig}-(a), and hence, we show $-\arg(T)$ rather than $\arg(T)$. Invisibility is observed at the shifted energy $\varep \approx 0.047$. (b-d) Modulus of the pressure field of the scattering solutions at the value of $\varep$ with unit transmission. 
	}
	\label{Ribbon_Scatt_AcNet_Fig} 
\end{figure*}

To illustrate invisibility of chiral mirror symmetric defects we performed two-dimensional numerical simulations of the Helmholtz equation (using COMSOL) in a structure reproducing the Kekulé ribbons with the same defects as in Fig.~\ref{Ribbon_Scatt_Fig}-(b-d). A monochromatic source is located on the left of the defect. Because the left and right moving modes have the same transverse profile, we apply the same method as one-dimensional waveguides to extract the reflection and transmission coefficients (see Supplemental material), from the acoustic pressure at two nodes before and two nodes after the defect. The invisibility is illustrated by the maximum transmission amplitude equal to unity at a $10^{-3}$ precision level and the zero phase in Fig.~\ref{Ribbon_Scatt_AcNet_Fig}-(a). Note that a constant (defect independent) energy shift is observed, which presumably comes from two-dimensional effects near the junctions. The results of  Fig.~\ref{Ribbon_Scatt_AcNet_Fig}, including the field profiles, reveal a remarkable agreement with the lattice model predictions (see Fig.~\ref{Ribbon_Scatt_Fig}).

\section{Conclusions and outlook}

We showed how a combination of chiral and mirror symmetry can lead to a protected unit transmission at zero energy. Although our results have mainly been discussed in the context of the Kekulé model, they are very general: our main theorem (before equation~\eqref{Theorem_eq}) will hold for any model with a single pair of propagating waves at zero energy and the commuting combination of chiral and mirror symmetries.
Moreover, such lattice models can be obtained in many classical waves systems~\cite{Freeney20,Zhang20,Wakao20}, and we numerically showed how to realize topological invisibility with acoustic waves in a network of tubes. 
The possibility of obtaining robust waveguiding in passive topological metamaterials should make it very appealing for applications and we hope that our findings will foster more studies in that direction.

\section*{Aknowledgements}

AC would like to thank T.~Torres for discussions on the topic. 
This project has received funding from the European Union's Horizon 2020 research and innovation programme under the Marie Sklodowska-Curie grant agreement No 843152. 
V.A. acknowledges financial support from the NoHENA project funded under the program Etoiles Montantes of the Region Pays de la Loire.

\section*{Author contributions}
A.~C. did the model analysis and numerical computations of the lattice model. 
L.-Y.~Z. performed the COMSOL simulations and V.~P. extracted the scattering coefficient from the COMSOL data. 
A.~C. wrote the manuscript. 
All authors initiated the project, discussed the results and implications, and commented on the manuscript at all stages.

\bibliographystyle{utphys}
\bibliography{Bibli}


%
%

\newpage 
\appendix
\begin{center}
\textbf{\large Supplemental material}
\end{center}
\vspace{-25pt}

\section{Mirror winding number}

Here, we briefly reproduce the derivation of the mirror winding numbers, but refer to~\cite{Kariyado17} for more details. We consider a molecular zigzag edge along the direction $\mathbf{a_\parallel} = 2\mathbf{a_1}+\mathbf{a_2}$, which is invariant under the mirror symmetry $\M$ about the axis $\mathbf{a_2}$. When the Bloch momentum $\mathbf{q}$ is parallel to $\mathbf{a_\perp}$, $\M$ commutes with the Bloch Hamiltonian $H(\mathbf{q})$. Hence, it can be split into a (one-dimensional) symmetric $H_S(q_\perp)$ and antisymmetric $H_A(q_\perp)$ part. Because $\M$ commutes with $\Gam$, both $H_S$ and $H_A$ are chiral. This means that a winding number can be computed for each part. For a molecular zigzag edge, a unit cell is chosen as in Fig.~\ref{Kekule_Fig}-(a). We write $|j\rangle$ with $j=1..6$ the vector with unit amplitude on site $j$ and zero elsewhere. Now, the symmetric sector is spanned by 
\be \label{Molecule_Sym_Subspace}
|1\rangle, \quad \frac1{\sqrt 2} (|2\rangle + |3\rangle), \quad \frac1{\sqrt 2} (|4\rangle + |6\rangle) , \quad |5\rangle, 
\ee 
while the antisymmetric sector is spanned by 
\be \label{Molecule_ASym_Subspace}
\frac1{\sqrt 2} (|2\rangle - |3\rangle), \quad \frac1{\sqrt 2} (|4\rangle - |6\rangle) . 
\ee 
In this basis, the two sub-Hamiltonian at $q_\parallel = 0$ have the chiral form 
\be 
H_{S/A}(q_\perp) = \bmat 0 & Q_{S/A}(q_\perp) \\ Q_{S/A}(q_\perp)^\dagger & 0 \emat , 
\ee 
with 
\bsub \bea 
Q_S(q_\perp) &=& \bmat s \sqrt 2 & t e^{2 i q_\perp} \\ s + t e^{-i q_\perp} & s \sqrt 2 \emat , \\
Q_A(q_\perp) &=& s - t e^{-i q_\perp}. 
\eea \esub 
An explicit computation of the winding numbers $\frac1{2i\pi} \int \mathrm{Tr}(Q^{-1} \p_{q_\perp} Q) \d q_\perp$ leads to 
\bsub \label{ZZ_MirrorWindingNumber} \bea 
(n_S, n_A) &=& (0,0) \qquad \text{if} \qquad s>t , \\
(n_S, n_A) &=& (1,-1) \qquad \text{if} \qquad s<t . 
\eea \esub 
We emphasize that nontrivial mirror winding numbers not only guarantee the existence of edge waves, but also a Dirac point at $\varep=0$ (gapless edge modes). This is because both winding numbers imply a zero energy solution localized near the edge. Moreover, the sign of the winding number tells us on which sublattice that solution is. 

\section{Dirac points protected by chiral and mirror symmetries}

When a two-dimensional system displays commuting chiral symmetry and a spatial symmetry, Dirac points can be protected by the symmetry combination. Typically, one can use chiral symmetry at high symmetry points to count the number of zero-modes for each spatial symmetry eigenvalue, as explained in details in~\cite{Koshino14}. In our work, the spatial symmetry is mirror symmetry. The simplest example of such Dirac point is provided by a simple chain (see Fig.~\ref{Principle_Fig}-(b)) with two sites per unit cell (one $\mathcal A$ and one $\mathcal B$). In this case, the Dirac point is simply that coming from the band folding of the simple chain, but the symmetry argument implies that it is stable under all symmetry preserving perturbation. To see this, we consider the mirror symmetry with axis on the right site of a unit cell. For a general value of $q$, that symmetry acts as 
\be 
M(q) = \bmat e^{iq} & 0 \\ 0 & 1 \emat, 
\ee 
such that the Bloch Hamiltonian satisfies 
\be 
M(q)^{\dagger} H(q) M(q) = H(-q). 
\ee 
At $q=0$, we see that both basis vectors, i.e. $|1\rangle$ and $|2\rangle$, belong to the same eigenspace of $M$ (of eigenvalue $1$). Hence, that space has 1 site on $\mathcal A$ and one on $\mathcal B$, and has no symmetry protected zero-mode. On the contrary, at $q=\pi$, $|1\rangle$ is anti-symmetric (eigenvalue $-1$) while $|2\rangle$ is symmetric (eigenvalue $1$). Hence, each sector has a single state belonging to a certain sublattice, which implies it can only have zero energy. This shows that the gap must close at $q=\pi$ for $\varep = 0$. 

In the core of the manuscript, we applied the same argument to a finite width ribbon. Although the argument is the same, the counting is more involved. We distinguish two types of sites: they either come in mirror pairs, or are on the symmetry axis. At $q=0$, the mirror symmetry has a block diagonal form, acting trivially on sites on-axis and exchanging mirror pairs off-axis. Notice that a supercell contains two symmetry axes. If $M$ is defined with respect to one of them, at $q=0$ sites on the other axis are also invariant under the action of $M$, because they would be mapped to sites on the same axis in the next supercell, which means they pick up a phase $e^{iq} = 1$. Thus, with an appropriate site labeling the mirror operator $M$ reads 
\be 
M = \bmat 1 & & & & & & & \\
& \ddots & & & & & & \\
& & 1 & & & & & \\
& & & 0 & 1 & & & \\ 
& & & 1 & 0 & & & \\ 
& & & & & \ddots & & \\
& & & & & & 0 & 1 \\
& & & & & & 1 & 0
\emat . 
\ee 
Using the eigenbasis of $M$, counting the difference of number of basis vector on each sublattice of the symmetric (resp. anti-symmetric) subspace gives us the number of symmetric (resp. anti-symmetric) zero modes. We can build basis vectors of the symmetric subspace as in equation~\eqref{Molecule_Sym_Subspace} with  
\be 
|j_\textrm{on axis}\rangle, \quad \text{and} \quad \frac1{\sqrt 2} (|j_\textrm{off axis} \rangle + |j'_\textrm{off axis} \rangle), 
\ee 
where $j'_\textrm{off axis}$ is the mirror symmetric of the site $j_\textrm{off axis}$. The anti-symmetric subspace is spanned by 
\be 
\quad \frac1{\sqrt 2} (|j_\textrm{off axis} \rangle - |j'_\textrm{off axis} \rangle), 
\ee
similarly to equation~\eqref{Molecule_ASym_Subspace}. 

Because the chiral and mirror symmetries commute, each sub-Hamiltonian is chiral, in other words, each basis vector belongs to a definite sublattice. To perform the counting, we decompose a ribbon supercell as $2N_y$ molecules and 4 extra sites to make the upper partially bearded edge (see Fig.~\ref{Kekule_Fig}-(b) or Fig.~\ref{Ribbon_Mmat_Fig}). Let us first consider a single molecule: it has 2 sites on the symmetry axis, and 2 pairs off-axis. That gives 4 symmetric basis vector with 2 on $\mathcal A$ and 2 on $\mathcal B$, and 2 anti-symmetric ones with 1 $\mathcal A$ and 1 on $\mathcal B$ (see equations \eqref{Molecule_Sym_Subspace} and \eqref{Molecule_ASym_Subspace}). We see that each molecule has an equal number of vector on both sublattice for each symmetry sector. The unbalance comes from the 4 extra sites of Fig.~\ref{Kekule_Fig}-(b). They consist in 2 sites on the axis and 1 pair off-axis, which leads to 2 symmetric basis vector on $\mathcal B$ and 1 on $\mathcal A$, and a single anti-symmetric one on $\mathcal A$. In total, the symmetric ribbon Hamiltonian is obtained with $4 N_y+1$ basis vector on $\mathcal A$ and $4 N_y+2$ on $\mathcal B$, while the anti-symmetric ribbon Hamiltonian is obtained with $2N_y+1$ basis vector on $\mathcal A$ and $2N_y$ on $\mathcal B$. Hence, as used in the core of the manuscript, there is one symmetric zero-mode on $\mathcal B$ and one anti-symmetric zero-mode on $\mathcal A$. 

Notice also that the same counting argument can be done for the other high symmetry point $q=\pi$. However, the fact that a supercell is not mirror symmetric itself leads to a different repartition between sublattices for each symmetry. Indeed, all sites on the second symmetry axis now pick up a minus sign when mirror symmetry is applied. Hence, these sites contribute to the anti-symmetric basis rather than the symmetric ones. A detailed counting shows that each symmetry sector is balanced, and thus, no zero energy solution is symmetry protected. This explains why Kekulé edge waves have a single Dirac point rather than one at each high symmetry point. This is an important point in this work, as the perfect transmission is protected only when there is a single pair of propagating modes at $\varep = 0$. 

\section{Transfer matrix formalism: Eigenmodes}

\begin{figure}[htp]
	\centering
	\includegraphics[width=\columnwidth]{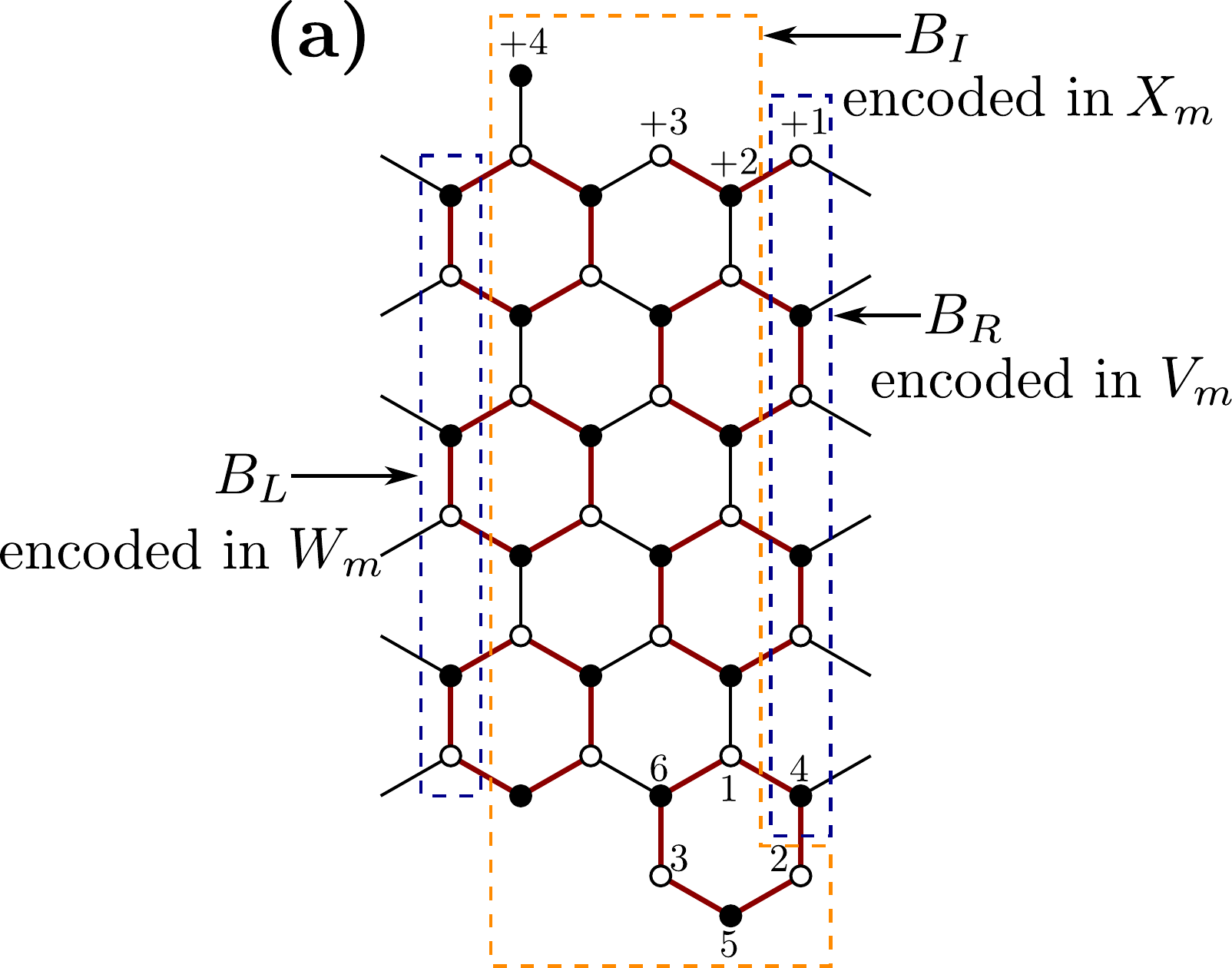}
	\caption{Supercell of a ribbon with mixed edges. 
	}
	\label{Ribbon_Mmat_Fig} 
\end{figure}

In this work, the scattering coefficients are obtained using a transfer matrix formalism, which we adapted from~\cite{Dwivedi16} to the present case. The starting point is to write the ribbon eigenvalue problem as 
\be \label{DC_General_eq}
J^\dagger  \Phi_{m-1} + H_{\rm sc}  \Phi_m + J  \Phi_{m+1} = \varep \Phi_m. 
\ee
In this equation, $\Phi_m$ is a column vector containing the amplitudes on all sites of the supercell $m$. Moreover, $J$ (resp. $J^\dagger$) is the matrix containing the hopping relating sites within a supercell to the next (resp. previous) supercell at $m+1$ (resp. $m-1$), and $H_{\rm sc}$ is the Hamiltonian of an isolated supercell. We now split the set of sites within a supercell in three subsets: $B_L$ (resp. $B_R$) contains sites connected to the supercell $m-1$ (rep. $m+1$), while $B_I$ contains sites that are only connected to other sites of the same supercell. This decomposition is illustrated in Fig.~\ref{Ribbon_Mmat_Fig}. Notice that periodicity implies that $B_L$ and $B_R$ have the same size, which is $2N_y$ for the ribbons we consider (see Fig.~\ref{Ribbon_Mmat_Fig}). We now write the solution $\Phi_m$ of \eq{DC_General_eq} in blocks: 
\be \label{DC_block_vec}
\Phi_m = \bmat (\phi_j)_{j\in B_R} \\ (\phi_j)_{j\in B_L} \\ (\phi_j)_{j\in B_I} \emat = \bmat V_m \\ W_m \\ X_m \emat . 
\ee
This block decomposition changes \eq{DC_General_eq} into 
\bsub \label{Block_Eig_eq} \bea
V_m &=& t \mathcal G_{vv} W_{m+1} + t \mathcal G_{vw} V_{m-1} , \qquad \label{Block_Eig_V} \\
W_m &=& t \mathcal G_{wv} W_{m+1} + t \mathcal G_{ww} V_{m-1} , \qquad \label{Block_Eig_W} \\
X_m &=& t \mathcal G_{xv} W_{m+1} + t \mathcal G_{xw} V_{m-1} , \qquad \label{Block_Eig_X}
\eea \esub
with $\mathcal G$ the supercell Green function $\mathcal G(\varep) = (\varep - H_{\rm sc})^{-1}$ written in block components 
\be \label{Block_Green}
\mathcal G(\varep) = \bmat \mathcal G_{vv} & \mathcal G_{vw} & \mathcal G_{vx} \\ \mathcal G_{wv} & \mathcal G_{ww} & \mathcal G_{wx} \\ \mathcal G_{xv} & \mathcal G_{xw} & \mathcal G_{xx} \emat . 
\ee
Since \eq{Block_Eig_X} gives us $X_m$ but does not involves $X_{m\pm 1}$, we can directly relate $W_m$ and $V_{m-1}$ to $W_{m+1}$ and $V_{m}$. This defines the transfer matrix: 
\be \label{DC_Mmat_def}
\bmat W_{m+1} \\ V_{m} \emat = M  \bmat W_{m} \\ V_{m-1} \emat , 
\ee
with  
\be \label{DCmat_general_prod}  
M = - \bmat t \mathcal G_{vv} & -\id_{2N_y} \\ t \mathcal G_{wv} & 0 \emat^{-1} \bmat 0 & t \mathcal G_{vw} \\ -\id_{2N_y} & t \mathcal G_{ww} \emat .  
\ee 
Eigenvectors of the transfer matrix are the modes of the ribbon. By construction, there are $4N_y$ of them. If the corresponding eigenvalue $\lam$ has a unit modulus, it is a purely propagating mode, otherwise it is evanescent. Moreover, all modes have a direction of propagation: if $|\lam | < 1$ it is moving to the right, and $|\lam | > 1$ it is moving to the left. If it is purely propagating ($|\lam |=1$), there are two options to determine its direction of propagation: one can compute the group velocity $v_g$, or one can add a small fictitious dissipation $\varep \to \varep + i \nu$ with $\nu>0$ and apply the previous criterion.

In order to have unambiguously defined scattering coefficients (see below), the last step is to properly normalize and fix the phase of the propagating modes $\varphi_\pm(\varep)$. We first normalize them by requiring that they carry a unit (up to the sign) conserved current (this is detailed in the next section). We then fix the phase by requiring a zero phase on a well-chosen site. Importantly, this site must be chosen so that the symmetry identities~\eqref{Chiral_Modes} and \eqref{Mirror_Modes} are valid. Since $\Gamma$ act on a site in $\mathcal B$ as multiplication by $1$, the chiral identity~\eqref{Chiral_Modes} requires this site ot belong to sublattice $\mathcal B$. Similarly, for the mirror identity~\eqref{Mirror_Modes} to hold, the phase-fixing site must be on the symmetry axis (where $\M$ acts as multiplication by $1$). These two constraints imply that the site must be labelled 5 within a given molecule (see Fig.~\ref{Kekule_Fig}-(a)). Then, any molecule within a supercell work adequately, and we chose the lowest one on the right (see Fig.~\ref{Ribbon_Mmat_Fig}). Notice that this phase fixing is possible because the amplitude of $\varphi_\pm(\varep)$ does not vanish on site $5$ (at least for $\varep$ near $0$). Interestingly, this requirement is what distinguish the different roles played by sublattice $\mathcal A$ and $\mathcal B$. Indeed, one could have tried to define $\Gamma = \mathrm{diag}(\mathbf{1}_{\mathcal A},- \mathbf{1}_{\mathcal B})$ instead of~\eqref{Chiral_Op}, and fix a zero phase on sublattice $\mathcal A$. However, sites both on $\mathcal A$ and the symmetry axis  are labeled 1 in Fig.~\ref{Kekule_Fig}-(a), and at $\varep=0$, the amplitudes of both $\varphi_S$ and $\varphi_A$ vanish on them (the former because it is on sublattice $\mathcal A$ and the latter because it is on the symmetry axis), and hence the same is true for $\varphi_\pm$ (which can be seen in Fig.~\ref{Kekule_Fig}-(b)).

\section{Transfer matrix formalism: conserved energy current}
\label{DC_Current_App}

A key point in \eq{DC_General_eq} is that $H_{\rm sc}$ is self-adjoint, and that the inter-supercell matrices relating $m\to m-1$, and $m\to m+1$ are adjoint. This guarantees that there is a conserved current along the ribbon. To see this, we take the (left) product of \eq{DC_General_eq} with $\Phi_m^\dagger$. Using the fact that $\Phi_m^\dagger (\varep - H_{\rm sc}) \Phi_m$ is real, we obtain 
\be
\Im\left( \Phi_m^\dagger J^\dagger \Phi_{m-1} \right) + \Im\left( \Phi_m^\dagger J \Phi_{m+1} \right) = 0 . 
\ee
This can now be written as 
\be
\mathcal J_{m+1} - \mathcal J_{m} = 0, 
\ee
with 
\be \label{Current_def}
\mathcal J_m = \Im\left( \Phi_{m-1}^\dagger J \Phi_{m} \right) . 
\ee
Therefore, $\mathcal J_m$ is a conserved current. Using the block form of $\Phi_m$, i.e. \eq{DC_block_vec}, this becomes 
\be \label{Current_Block}
\mathcal J_m = t \Im\left( V_{m-1}^\dagger W_{m} \right) . 
\ee
This defines a symplectic structure: the fact that $\mathcal J_m$ is independent of $m$ implies that $M \in \mathrm{Sp}(4N_y,\mathbb R)$~\cite{Dwivedi16}. More generally, the current conservation stays valid for any non-periodic ribbon of similar shape. Indeed, if the hoppings depend on the supercell, \eq{DC_General_eq} is changed into 
\be \label{DC_Inhomo_eq}
J_{m-1}^\dagger \cdot \Phi_{m-1} + H_m^{\rm sc} \cdot \Phi_m + J_m \cdot \Phi_{m+1} = \varep \Phi_m . 
\ee
Notice that unitarity imposes that there is only one set of hopping matrices $J_m$ and not a left and right one. Following the same steps as before, we obtain he conserved current 
\be
\mathcal J_m = \Im\left( \Phi_{m-1}^\dagger J_m \Phi_{m} \right) , 
\ee 
which generalizes \eq{Current_def}. In particular, the current \eqref{Current_def} will be conserved across any type of defect or disorder.

\section{Transfer matrix formalism: scattering matrix}

\begin{figure}[htp]
	\centering
	\includegraphics[width=\columnwidth]{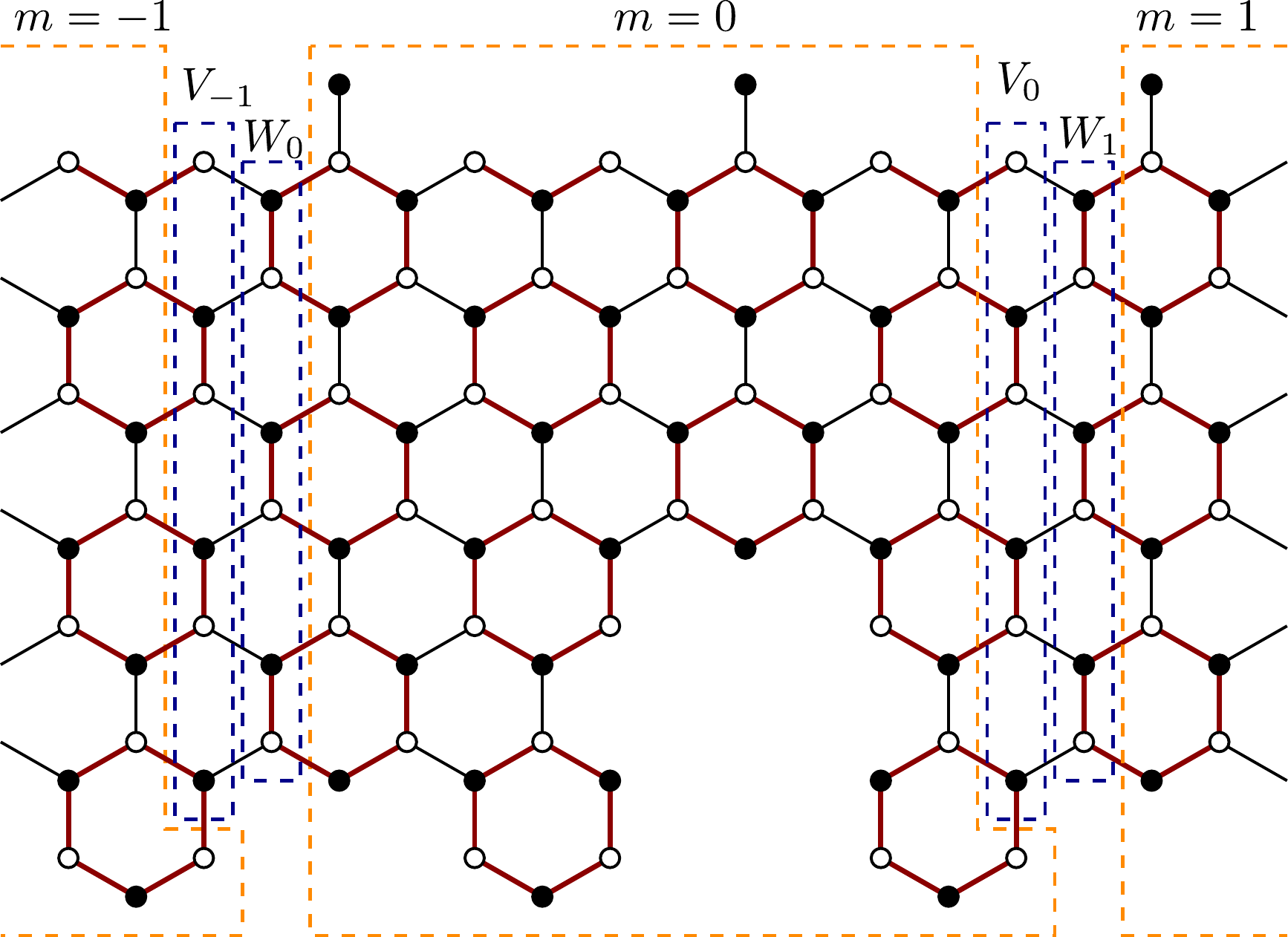}
	\caption{Kekulé ribbon with a simple defect made of a missing molecule. Decomposition of the defect supercell and neighboring supercells are emphasized. 
	}
	\label{DCmat_Scatt_Fig} 
\end{figure}

To compute the scattering matrix over a defect, we define the supercell $m=0$ so that the defect is entirely contained within it. This means that for $m \neq 0$ the eigenvalue equation is given by \eq{DC_General_eq}, and for $m=0$ it has the same form with $H_{\rm sc}$ replaced by a defect Hamiltonian $H_{\rm d}$. This is illustrated in Fig.~\ref{DCmat_Scatt_Fig}. A scattering solution can then be written as: 
\bea \label{Defect_scatt_sol} 
\bmat W_{m} \\ V_{m-1} \emat &\underset{m \leqslant 0}{=}& \sum_{j=1}^{2N_y} a_j (\lam_j^{+})^{m} \varphi_j^{+} + b_j (\lam_j^{-})^{m} \varphi_j^{-} , \nonumber \\ 
&\underset{m > 0}{=}& \sum_{j=1}^{2N_y} c_j (\lam_j^{+})^{m-1+n_d} \varphi_j^{+} , 
\eea 
where $\varphi_j^{\pm}$ are the eigenvectors of $M$ with eigenvalues $\lam_j^{\pm}$ propagating to the right ($+$) or left ($-$). $a_j$ are the incident amplitudes, related to the outgoing amplitudes by the reflection matrix $(b_j) = \hat{R}(a_j)$ and transmission matrix $(c_j) = \hat{T}(a_j)$. 
Notice that we also introduce $n_d$, the size of the defect supercell counted in number of ribbon supercells that would have the same horizontal length (for instance $n_d=2$ in Fig.~\ref{DCmat_Scatt_Fig}), ensuring that $T=1$ in the absence of defect. Following the same steps that lead to \eq{Block_Eig_eq}, we can relate the amplitudes on both sides of the defect using the defect Green function $\mathcal G^\d(\varep) = (\varep - H_{\d})^{-1}$ and its block components as in \eq{Block_Green}: 
\bsub \bea
V_{0} &=& t \mathcal G^{\rm d}_{vv}  W_1 + t \mathcal G^{\rm d}_{vw}  V_{-1} , \\
W_{0} &=& t \mathcal G^{\rm d}_{wv}  W_1 + t \mathcal G^{\rm d}_{ww}  V_{-1} . 
\eea \esub
Using the block form of the scattering solution defined in~\eqref{Defect_scatt_sol}, we obtain the set of equations 
\bsub \bea
\hat{V}^{+}  (D^{+})^{n_d}  \hat{T} &=& t \mathcal G^{\rm d}_{vv}  \hat{W}^{+}  (D^{+})^{n_d}  \hat{T}  \nonumber \\
&& + t \mathcal G^{\rm d}_{vw}  \hat{V}^{+} + t \mathcal G^{\rm d}_{vw}  \hat{V}^{-}  \hat{R} , \qquad \\
\hat{W}^{+} + \hat{W}^{-}  \hat{R} &=& t \mathcal G^{\rm d}_{wv}  \hat{W}^{+}  (D^{+})^{n_d}  \hat{T} \nonumber \\ 
&& + t \mathcal G^{\rm d}_{ww}  \hat{V}^{+} + t \mathcal G^{\rm d}_{ww}  \hat{V}^{-}  \hat{R} . \qquad 
\eea \esub
In these equations, we defined $\hat{W}^\pm$ (resp. $\hat{V}^\pm$) two square matrices made by columns with the $W$-components (resp. $V$-components) of the modes $(\varphi_j^{\pm})_{j=1..2N_y}$. We also defined the diagonal matrices $D^\pm$ with the eigenvalues $(\lam_j^\pm)_{j=1..2N_y}$ in the diagonal. This system can be written 
\be
\hat{\alpha}  \bmat \hat{R} \\ \hat{T} \emat = \hat{\beta} , 
\ee
with the block matrices 
\bea
\hat{\alpha} &=& \bmat -t \mathcal G^{\rm d}_{vw}  \hat{V}^{-} & (\hat{V}^{+} - t \mathcal G^{\rm d}_{vv}  \hat{W}^{+})  (D^{+})^{n_d}\\
W^{-} - t \mathcal G^{\rm d}_{ww}  \hat{V}^{-} & -t \mathcal G^{\rm d}_{wv}  \hat{W}^{+}  (D^{+})^{n_d}  \emat , \nonumber \\
\hat{\beta} &=& \bmat t \mathcal G^{\rm d}_{vw}  \hat{V}^{+} \\ t \mathcal G^{\rm d}_{ww}  \hat{V}^{+} - \hat{W}^+ \emat , 
\eea 
which we solve to obtain $\hat{R}$ and $\hat{T}$. We also normalized all propagating modes so that they carry a unit conserved current as defined in \eq{Current_Block}. This gives us relations between the reflection and transmission coefficients restricted to purely propagating modes. For the energy range of edge waves, there are only two such coefficients that we call $R(\varep)$ and $T(\varep)$. Current conservation leads to 
\be 
|T|^2 + |R|^2 = 1. 
\ee

\section{Defect topological indices}
\label{Lemma_App}

\begin{figure}[htp]
	\centering
	\includegraphics[width=\columnwidth]{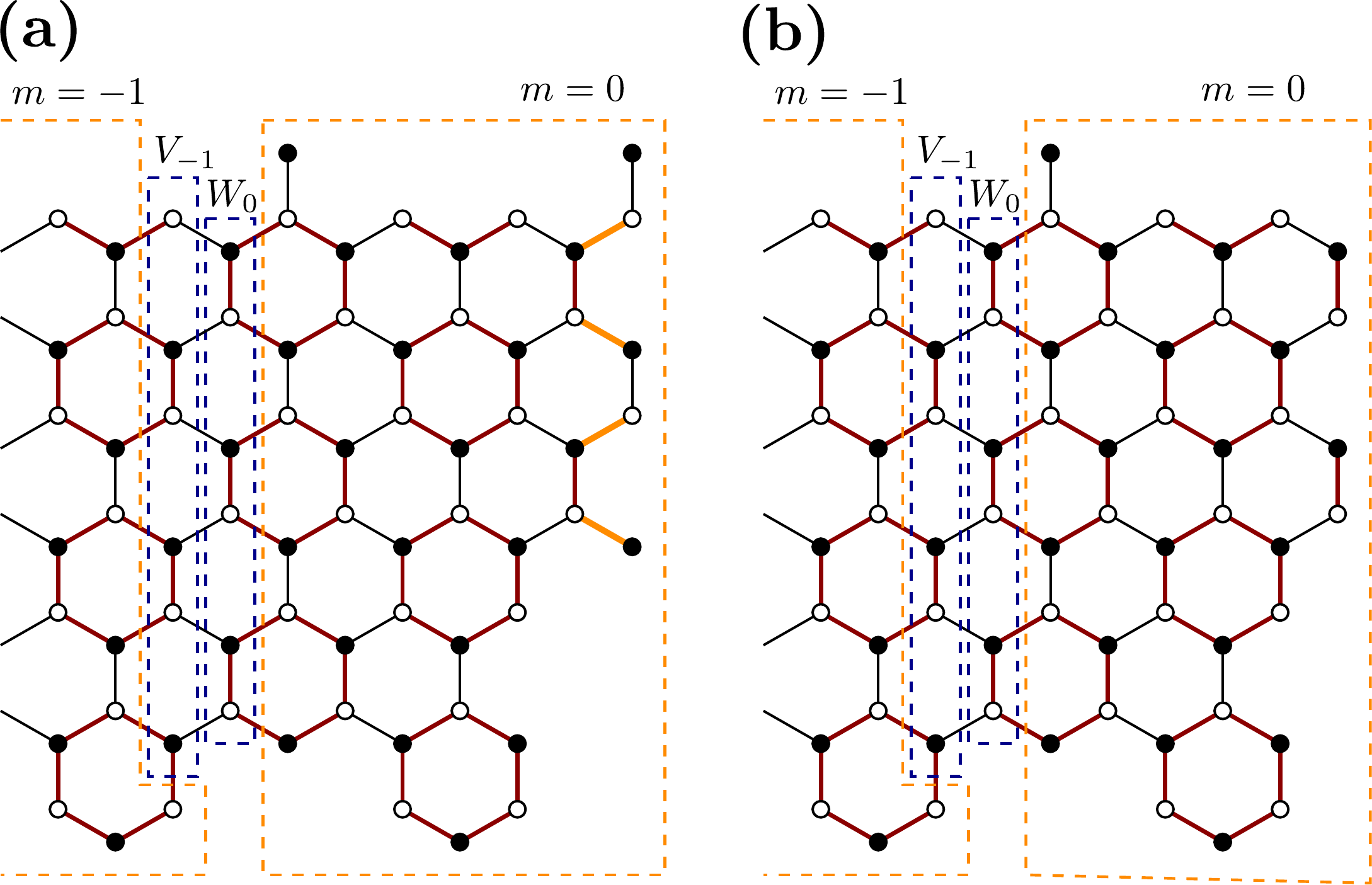}
	\caption{One-port scattering problem obtained by symmetry reduction of the two-port scattering problem on the simple defect of Fig.~\ref{DCmat_Scatt_Fig}. (a) symmetric subproblem (b) anti-symmetric subproblem. 
	}
	\label{SymAsym_Scatt_Fig} 
\end{figure}

In the main text, we saw that finding the values of the zero energy reflection and transmission coefficients boils down to finding on which sublattice the solution of the one-port scattering problems of each symmetry sector is supported (see Fig~\ref{SymAsym_Scatt_Fig} for an illustration of the two one-port scattering problems obtained by symmetry reduction). It turns out that the only information needed to conclude is given by the topological indices of the defect (as defined in equation~\eqref{TopoIndices}). Here we formulate a general conjecture, which allows us to treat defects of any indices: 
\bigskip 

\noindent {\bf Conjecture:} 
The scattering solution of the one-port problem corresponding to the symmetric (resp. anti-symmetric) sector has support on sublattice $\mathcal B$ (resp. $\mathcal A$) if and only if $\Delta_S \geq 0$ (resp. $\Delta_A \leq 0$). 
\bigskip 

A direct implication of that conjecture is that a topologically trivial defect, i.e. such that $(\Delta_S,\Delta_A) = (0,0)$, implies that $\Phi_S$ has support on sublattice $\mathcal B$ and $\Phi_A$ on $\mathcal A$, just as in the absence of defect,  
which is the result we used to prove our main theorem and equation~\eqref{Theorem_eq}. 

To support that conjecture, we solved the one-port scattering problem using the transfer matrix method for a variety of defects, and compared the values of $R_S(0)$ and $R_A(0)$ with the topological indices of the defects. The results are shown in Table~\ref{Defect_Table_Fig} and all agree with the general conjecture. As a final remark, we point out that the topological indices have an additive property: if we build a defect by taking out a first (symmetric) defect, and then a second one, the indices of the total defect are obtained by summing the indices of the first and second defects. This greatly simplifies the characterization of large defects, as they can be seen as resulting from the addition of smaller defects. 

\begin{table*}[htp]
	\centering
	\includegraphics[width=\textwidth]{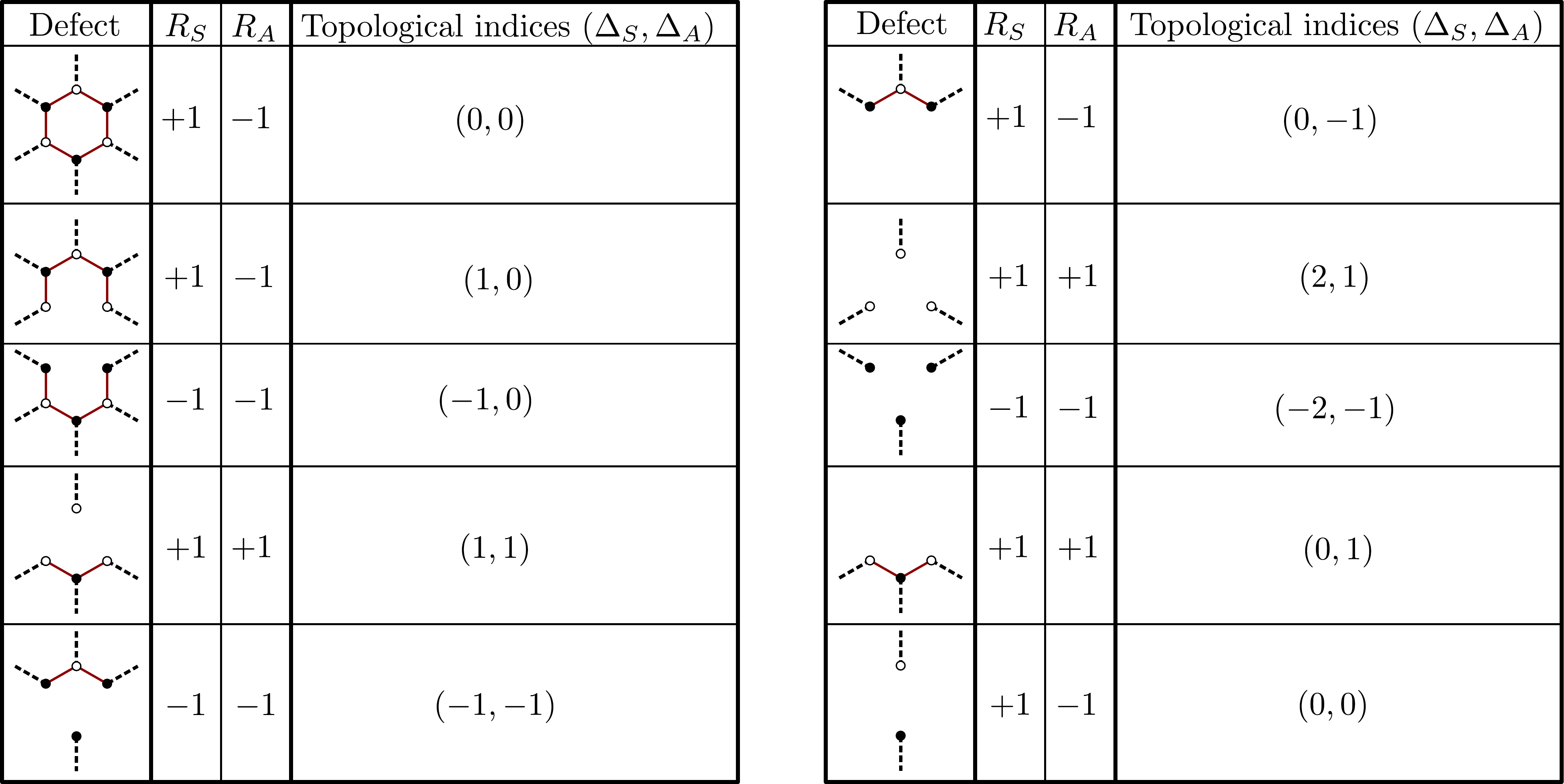}
	\caption{Mirror symmetric defects and their corresponding topological indices. The defects pictures show what have been cut out, as the example of Fig.~\ref{Symmetry_Fig}-(c). 
	}
	\label{Defect_Table_Fig} 
\end{table*}

\section{Extracting scattering coefficients}
\label{Num_Exp_Sec}

\begin{figure*}[htp]
	\centering
	\includegraphics[width=\textwidth]{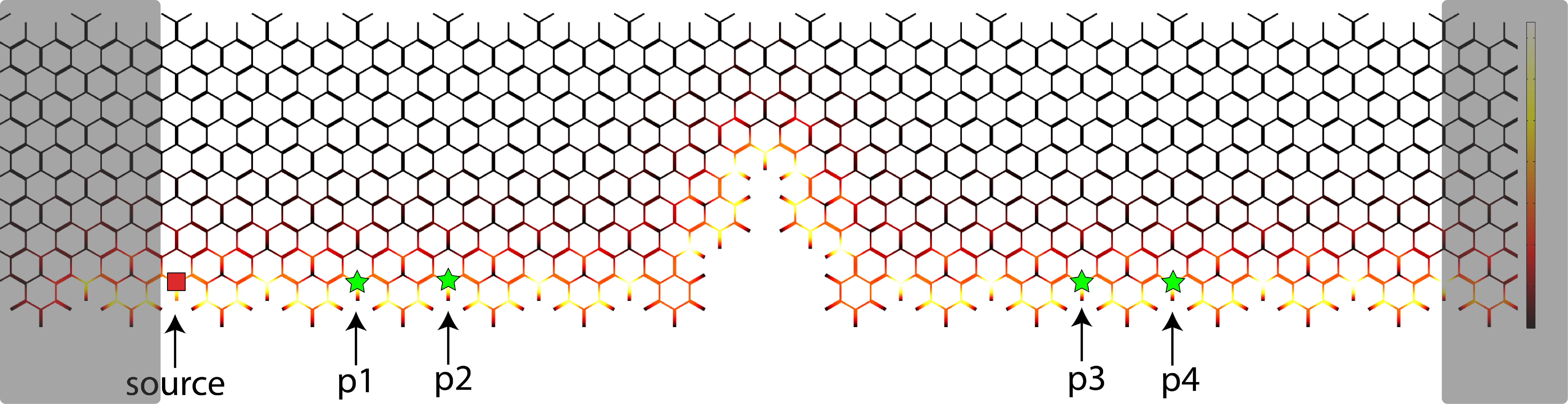}
	\caption{Numerical scattering experiment. We solve the Helmholtz equation in the acoustic network, with a monochromatic source on one intersection (red square), and obtain the pressure amplitude (for varying energies) on 4 different intersections (green stars): $p_1$, $p_2$, $p_3$ and $p_4$. In this figure the site of $p_1$ is separated by its mirror symmetric $p_4$ by $n_0 = 9$. We also added dissipation on the far left and right sides (grey area) to mitigate resonances with the whole structure. 
	}
	\label{Scatt_NExp_Fig} 
\end{figure*}

We show how to extract scattering coefficients from numerical simulations of a problem with source, as was done to produce Fig.~\ref{Ribbon_Scatt_AcNet_Fig}. The method is adapted from~\cite{Abom91} to the present context. The numerical simulation provides us with a solution $\Phi_s$ at fixed energy $\varep = \cos(kL)$ and a source located at a chosen site (network node shown in Fig.~\ref{Scatt_NExp_Fig}). The energy is chosen in the range around zero such that only edge waves propagate. On each side of the defect, sufficiently far that near field effects can be neglected, the solution is a superposition of traveling waves: 
\bsub \label{Source_Sol} \bea 
\Phi_s &=& A_L \varphi_+ + B_L \varphi_- \quad \text{(left side)}, \\
\Phi_s &=& A_R \varphi_+ + B_R \varphi_- \quad \text{(right side)}.  
\eea \esub 
We now proceed in two steps: first we obtain the above decomposition from pressure values on well chosen nodes, and second we extract the scattering coefficients. In the simulations, we take the values of $\Phi_s$ on two sites for each side of the defect (see Fig.~\ref{Scatt_NExp_Fig}), called $p_1$, $p_2$ on the left and $p_3$, $p_4$ on the right. Moreover, we chose $p_2$ (resp. $p_4$) to be on the same site as $p_1$ (resp. $p_3$) in the next supercell to the right. This means that if we call $\varphi_\pm^{(j)}$ the mode amplitude on the site $j$ where $p_1$ (resp. $p_3$) is taken, then the mode amplitude where $p_2$ (resp. $p_4$) is taken is $e^{\pm iq} \varphi_\pm^{(j)}$. From equation~\eqref{Source_Sol}, we have explicitly 
\bsub \label{pressure_values} \bea 
p_1 &=& A_L \varphi_+^{(j)} + B_L \varphi_-^{(j)} , \\
p_2 &=& A_L e^{iq} \varphi_+^{(j)} + B_L e^{-iq} \varphi_-^{(j)} , 
\eea \esub 
and similarly for $p_3$ and $p_4$. We now need to relate the mode amplitude of the left and right moving waves on the chosen site $j$. For this we exploit the mirror symmetry of the problem, and the relation~\eqref{Mirror_Modes}. Calling $n_0$ the number of supercell separating the site $j$ from its mirror symmetric, we have $\varphi_-^{(j)} = e^{i n_0 q} \varphi_+^{(j)}$. Now equations~\eqref{pressure_values} can be rewritten in a matrix form 
\be 
\bmat p_1 \\ p_2 \emat = \varphi_+^{(j)} \bmat 1 & e^{in_0 q} \\ e^{iq} & e^{i(n_0-1)q} \emat \bmat A_L \\ B_L \emat , 
\ee 
which we invert to get 
\be \label{PressureToAmp}
\bmat A_L \\ B_L \emat = \frac{i}{2 \sin(q) \varphi_+^{(j)}} \bmat e^{-i q} & -1 \\ - e^{i(1-n_0)q} & e^{-i n_0 q} \emat \bmat p_1 \\ p_2 \emat. 
\ee 
To obtain $A_R$ and $B_R$ as function of $p_3$ and $p_4$ we notice that mirror symmetric swaps the roles of $(A_L,B_L)$ with $(B_R, A_R)$, and $(p_1, p_2)$ with $(p_4, p_3)$. Hence, 
\be \label{PressureToAmp2}
\bmat B_R \\ A_R \emat = \frac{i}{2 \sin(q) \varphi_+^{(j)}} \bmat e^{-i q} & -1 \\ - e^{i(1-n_0)q} & e^{-i n_0 q} \emat \bmat p_4 \\ p_3 \emat. 
\ee 
Importantly, equations~\eqref{PressureToAmp} and \eqref{PressureToAmp2} require the value of $q$ at the chosen energy $\varep$. For this, we also numerically solved (finite elements in COMSOL) the dispersion relation for a ribbon without defect, and use it to obtain $q(\varep)$. 

For the second step, we decompose $\Phi_s$ on a basis of scattering solutions. A scattering solution with an incident wave on the left has a (far field) decomposition as written in equation~\ref{Scatt_as}. To emphasize that the incident wave comes from the left, we call this scattering solution $\Phi_L$ rather that $\Phi$ in the core of the manuscript. We now use a convenient property of mirror symmetric scattering problems, namely that the reflection and transmission coefficients are identical whether the incident wave comes from the left or right. Hence, the scattering solution $\Phi_R$ with incident wave on the right decomposes as 
\bsub \label{Scatt_as_R} \bea 
\Phi_R &=& T \varphi_- \quad \text{(left side)}, \\
\Phi_R &=& R \varphi_+ + \varphi_- \quad \text{(right side)}.  
\eea \esub 
Now, comparing the decomposition of $\Phi_s$ (equation~\eqref{Source_Sol}) with that of $\Phi_L$ (equation~\eqref{Scatt_as}) and $\Phi_R$ (equation~\eqref{Scatt_as_R}), we see that 
\be 
\Phi_s = A_L \Phi_L + B_R \Phi_R , 
\ee 
from which we deduce 
\bsub \bea 
B_L &=& A_L R + B_R T , \\
A_R &=& B_R R + A_L T . 
\eea \esub 
Inverting the above system for $R$ and $T$ leads to the expressions for the scattering coefficients 
\bsub \bea 
R &=& \frac{A_L B_L - A_R B_R}{A_L^2 - B_R^2} , \\
T &=& \frac{A_L A_R - B_R B_L}{A_L^2 - B_R^2} . 
\eea \esub 
Lastly, to obtain the (approximate) scattering solutions displaying perfect transmission shown in Fig.~\ref{Ribbon_Scatt_AcNet_Fig}-(b-d), we tuned the losses in the dissipative layers to minimize the coefficient $B_R$.

\end{document}